\documentclass[10pt,conference]{IEEEtran}
\IEEEoverridecommandlockouts
\usepackage{cite}
\usepackage{amsmath,amssymb,amsfonts}
\usepackage{algorithmic}
\usepackage{algorithm}
\usepackage{graphicx}
\usepackage[caption=false,font=footnotesize,labelfont=rm,textfont=rm]{subfig}
\usepackage{textcomp}
\usepackage{xcolor}
\usepackage{booktabs} % 在导言区加入这个宏包
\def\BibTeX{{\rm B\kern-.05em{\sc i\kern-.025em b}\kern-.08em
    T\kern-.1667em\lower.7ex\hbox{E}\kern-.125emX}}
\usepackage[left=1.62cm,right=1.62cm,top=1.9cm]{geometry}
\DeclareRobustCommand*{\IEEEauthorrefmark}[1]{%
	\raisebox{0pt}[0pt][0pt]{\textsuperscript{\footnotesize\ensuremath{#1}}}}
\begin{document}
\title{Energy Efficiency Optimization for Movable Antenna-Aided Communication Systems
}

\author{
	\IEEEauthorblockN{
		Jingze Ding\IEEEauthorrefmark{1}$^\ast$, 
		Zijian Zhou\IEEEauthorrefmark{2}, 
		Yuping Zhao\IEEEauthorrefmark{1},
		and Bingli Jiao\IEEEauthorrefmark{3}\IEEEauthorrefmark{1} 
		} 
	\IEEEauthorblockA{\IEEEauthorrefmark{1}School of Electronics, Peking University, Beijing 100871, China}
	\IEEEauthorblockA{\IEEEauthorrefmark{2}School of Science and Engineering, The Chinese University of Hong Kong, Shenzhen, Guangdong 518172, China}
	\IEEEauthorblockA{\IEEEauthorrefmark{3}School of Computing and Artificial Intelligence, Fuyao University of Science and Technology, Fuzhou, Fujian 350109, China\\
		$^*$Corresponding author: Jingze Ding (djz@stu.pku.edu.cn)\\
	Email: djz@stu.pku.edu.cn; zijianzhou@link.cuhk.edu.cn; yuping.zhao@pku.edu.cn; jiaobl@pku.edu.cn}
}

\maketitle

\begin{abstract}
This paper investigates the energy efficiency optimization for movable antenna (MA) systems by considering the time delay and energy consumption introduced by MA movement. We first derive the upper bound on energy efficiency for a single-user downlink communication system, where the user is equipped with a single MA. Then, the energy efficiency maximization problem is formulated to optimize the MA position, and an efficient algorithm based on successive convex approximation is proposed to solve this non-convex optimization problem. Simulation results show that, despite the overhead caused by MA movement, the MA system can still improve the energy efficiency compared to the conventional fixed-position antenna (FPA) system.
\end{abstract}


\section{Introduction}
Next-generation wireless communication systems require increasingly higher capacity to meet the demands of modern applications. Multiple-input multiple-output technology has emerged as a key solution by exploiting spatial degrees of freedom (DoFs) for both diversity and multiplexing gains. However, the antenna schemes implemented in existing systems are predominantly fixed-position antenna (FPA), which limits the exploitation of spatial DoFs available in the continuous spatial domain.

To harness the continuous spatial DoFs in wireless channels, movable antenna (MA) technology \cite{MA1} has recently been proposed to overcome the inherent constraints of discrete antenna placement in conventional FPA systems. Specifically, the MA is connected to the radio frequency (RF) chain via a flexible cable and it can be moved in a given region with the aid of a driver to offer remarkable capabilities in signal power enhancement, interference mitigation, and flexible beamforming \cite{MA2,near1}. Given the aforementioned advantages, the MA technology is widely used to improve the performance of existing wireless communication and sensing systems, such as full-duplex communications \cite{MAFD1,MAFD2,MAFD3}, satellite communications \cite{MAsate1,MAsate2}, integrated sensing and communication (ISAC) \cite{MAISAC1,MAISAC2}, and so on. 

The majority of existing studies concentrate on exploiting the flexible MA movement to reconfigure the channel, thereby maximizing the achievable rate or minimizing the transmit power. However, the MA movement introduces additional time and energy consumption. On the one hand, the time consumption reduces the time available for data transmission within a transmission block, thereby leading to a decrease in the achievable throughput \cite{MAtime}. On the other hand, the energy consumption requires terminals equipped with MAs to reserve sufficient energy for MA movement, which reduces the available energy for data transmission in energy-constrained systems \cite{MAenergy}. Therefore, these two types of consumption are non-negligible and need to be considered in the design of MA systems. Currently, there has been no prior work on optimizing MA systems by simultaneously considering the time delay and energy consumption introduced by MA movement. As such, this paper investigates the energy efficiency optimization for MA systems. First, we derive the upper bound of energy efficiency for an MA-aided single-user downlink communication system based on the time and energy consumption of MA movement. Next, we formulate the energy efficiency maximization problem to optimize the MA position and propose an effective algorithm based on successive convex approximation (SCA) to solve this non-convex optimization problem. Finally, we present simulation results to demonstrate that the MA system improves energy efficiency compared to the conventional FPA system, even in the presence of additional MA movement overhead.
\section{System Model and Problem Formulation}\label{2}
\subsection{System Model}\label{system_model}
\begin{figure}[!t]
	\centering
	\includegraphics[width=1\linewidth]{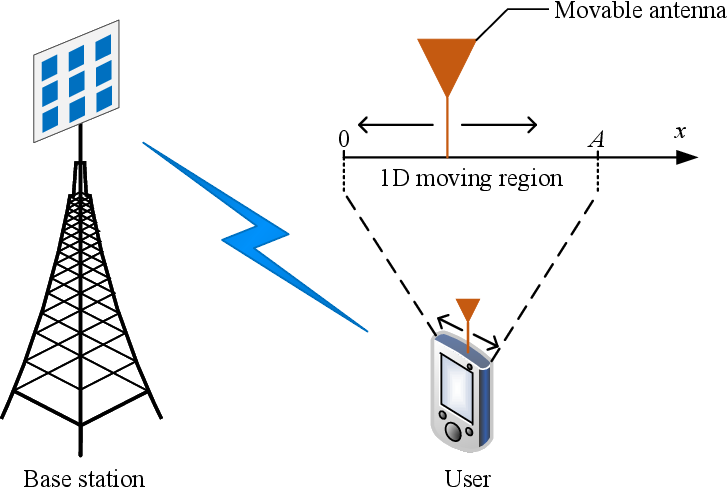}
	\caption{Illustration of the proposed MA-aided downlink communication system.}
	\label{sysmodel}
\end{figure}
As shown in Fig. \ref{sysmodel}, we consider a multiple-input single-output downlink communication system, where the base station (BS) is equipped with $N$ FPAs to serve a single-MA user. The MA is connected to the RF chain through a flexible cable such that it can move along a linear array without restraint. A one-dimensional (1D) local coordinate system in Fig. \ref{sysmodel} is established to describe the MA position, which is denoted as $x \in \mathcal{A}$, where $\mathcal{A}$ is the given 1D moving region with interval $\left[0,A \right]$. 

According to the field-response channel model \cite{MA2}, we consider narrow-band channels with slow fading and focus on one quasi-static fading block. The channel vector of MA systems is determined by the propagation environment and the MA position. Let $L$ denote the number of channel paths from the BS to the user. Define ${\vartheta _{l}} = \sin {\theta _{l}}\cos {\phi _{l}}$ ($1 \le l \le L$) as the virtual angle of arrival (AoA) of the $l$-th receive path at the user, where $\theta _{l}$ and $\phi _{l}$ denote the elevation and azimuth angles, respectively. The channel vector from the BS to the user can be expressed as 
\begin{equation}\label{channel}
	{\mathbf{h} } = \mathbf{G}^H{\mathbf{f}}\left( {{x}} \right),
\end{equation}
where $\mathbf{G} \in \mathbb{C}^{L \times N}$ is the path-response matrix representing the propagation environment. The entry in the $l$-th row and $n$-th ($1 \le n \le N$) column of $\mathbf{G}$ represents the response coefficient of the $l$-th channel path from the $n$-th antenna element at the BS to the reference point of $\mathcal{A}$. In addition, ${\mathbf{f}}\left( {{x}} \right) \in \mathbb{C}^{L \times 1}$ is the field-response vector representing the phase differences of $L$ channel paths determined by MA position, which is given by
\begin{equation}
	{\mathbf{f}}\left( {{x}} \right) = {\left[ {{e^{\mathrm{j}\frac{{2\pi}}{\lambda }{x}{\vartheta _{1}}}}, \ldots ,{e^{\mathrm{j}\frac{{2\pi}}{\lambda }{x}{\vartheta _{{L}}}}}} \right]^T},
\end{equation}
where $\lambda$ is the carrier wavelength.

Let $\mathbf{w}$ denote the transmit beamformer for the user. The received signal at the user can be expressed as
\begin{equation}
	{y} = \mathbf{h}^H{{\mathbf{w}}{s}}  + {n},
\end{equation}
where ${s}$ is the downlink information for the user with normalized power and ${n} \sim \mathcal{CN}\left( {0,\sigma ^2} \right)$ is the corresponding additive white Gaussian noise with zero mean and power $\sigma^2$. Then, the signal-to-noise ratio (SNR) of the user can be written as
\begin{equation}\label{SNR}
	{\Gamma} = \frac{{{{\left| {\mathbf{h}^H{\mathbf{w}}} \right|}^2}}}{\sigma^2}.
\end{equation}
\subsection{MA Energy Consumption Model and Problem Formulation}
The total energy consumption of MA systems consists of two components. The first one is the movement-related energy, which arises from the driver controlling the MA to change its position. The other component is the communication-related energy, which is due to the data transmission. For the movement-related energy, the energy consumption required for the considered 1D MA movement of the user in Section \ref{system_model} can be expressed as
\begin{equation}
	{E_{\mathrm{MA}}} = E_0\left| {x - x^0} \right|,
\end{equation}
where $E_0$ is the energy consumption rate of MA movement in Joule per meter (J/m), which represents the energy consumption for moving the MA by a unit distance. $x^0 \in \mathcal{A}$ is the initial MA position. We assume that the MA moves with a constant speed, and thus $E_0$ is also constant.

During the movement of the MA, the channel undergoes severe fluctuations. If data transmission occurs during this period, it may lead to significant degradation in the quality of service, thereby reducing energy efficiency. As a result, it is assumed that only the basic signaling overhead is maintained, while all data transmission is suspended during the MA movement. The slow-varying channel allows the repositioning of the MA to a location with more favorable channel condition at the beginning of each transmission block. Then, the user can receive the data transmitted from the BS with the assistance of fixed MA. Hence, a transmission block with time duration $T$ can be divided into two phases: 1) the movement phase and 2) the communication phase. As such, the total energy consumption for the user can be written as 
\begin{equation}\label{E_k}
	{E} =  {E_0\left| {{x} - x^0} \right|} +  {\left\| {{\mathbf{w}}} \right\|_2^2\left( {T -  {\frac{{\left| {{x} - x^0} \right|}}{{{v}}}}} \right)},
\end{equation}
where $v$ is the MA moving speed. On the other hand, during the communication phase, the achievable throughput in bits per Hertz (bits/Hz) of the user can be expressed as
\begin{equation}\label{R_k}
	{R} = \left( {T -  {\frac{{\left| {{x} - x^0} \right|}}{{{v}}}}} \right){\log _2}\left( {1 + {\Gamma}} \right).
\end{equation}
With \eqref{E_k} and \eqref{R_k}, the energy efficiency of the user can thus be expressed as
\begin{equation}
	EE=\frac{R}{E} = \frac{{\left( {T - \frac{{\left| {x - {x^0}} \right|}}{v}} \right){{\log }_2}\left( {1 + \frac{{{{\left| {{\mathbf{h}^H}\mathbf{w}} \right|}^2}}}{{{\sigma ^2}}}} \right)}}{{E_0\left| {x - {x^0}} \right| + \left\| \mathbf{w} \right\|_2^2\left( {T - \frac{{\left| {x - {x^0}} \right|}}{v}} \right)}}.
\end{equation}

In this paper, we aim to maximize the energy efficiency for the user by
jointly optimizing beamformer $\mathbf{w}$ and MA position $x$. The corresponding problem is formulated as
\begin{align}\label{max_single}
	& \mathop {\mathrm{max} }\limits_{\mathbf{w},x} \quad EE \\
	&\mathrm{s.t.} \quad \mathrm{C1}: {{\left( {T - \frac{{\left| {x - {x^0}} \right|}}{v}} \right){{\log }_2}\left( {1 + \frac{{{{\left| {{\mathbf{h}^H}\mathbf{w}} \right|}^2}}}{{{\sigma ^2}}}} \right)}} \ge R_\mathrm{TH},\nonumber\\
	&\hspace{2.3em} \mathrm{C2}: \left\| \mathbf{w} \right\|_2^2 \le P_\mathrm{t},\nonumber\\
	&\hspace{2.3em} \mathrm{C3}: x \in \mathcal{A}.\nonumber
\end{align}
where constraint C1 guarantees minimum throughput requirement $R_\mathrm{TH}$ for the user, constraint C2 indicates that the total transmit power of the BS should not exceed the maximum value, $P_\mathrm{t}$, and constraint C3 confines the moving region of the MA.  Note that problem \eqref{max_single} is a non-convex optimization problem with coupled variables, and thus we develop an efficient algorithm to obtain suboptimal solutions for this problem in the next section. 
\section{Proposed Solution}\label{3}
To gain essential insights into the MA position design and highlight the energy efficiency gains from MA movement, we use the maximum ratio combining (MRC) beamformer, i.e., $\mathbf{w} = \sqrt {{P_\mathrm{t}}} \frac{\mathbf{h}\left(x \right) }{{{{\left\| \mathbf{h}\left(x \right) \right\|}_2}}}$, which is a simple and effective combining method commonly used in single-user scenarios. Therefore, problem \eqref{max_single} can be formulated as
\begin{align}\label{max_single1}
	& \mathop {\mathrm{max} }\limits_{x} \quad \frac{{\left( {T - \frac{{\left| {x - {x^0}} \right|}}{v}} \right){{\log }_2}\left( {1 + \frac{{{P_\mathrm{t}}\left\| \mathbf{h}\left(x \right) \right\|_2^2}}{{{\sigma ^2}}}} \right)}}{{E_0\left| {x - {x^0}} \right| + {P_\mathrm{t}}\left( {T - \frac{{\left| {x - {x^0}} \right|}}{v}} \right)}} \\
	&\mathrm{s.t.} \quad \mathrm{C3}, \nonumber\\
	&\hspace{2.3em} \mathrm{C4}: {{\left( {T - \frac{{\left| {x - {x^0}} \right|}}{v}} \right){{\log }_2}\left( {1 + \frac{{{P_\mathrm{t}}\left\| \mathbf{h}\left(x \right) \right\|_2^2}}{{{\sigma ^2}}}} \right)}} \ge R_\mathrm{TH}.\nonumber
\end{align}
The upper bound of problem \eqref{max_single1} can be obtained by the following theorem.
\newtheorem{theorem}{Theorem}
\begin{theorem} \label{theorem1}
	The upper bound on the energy efficiency in problem \eqref{max_single1} is achieved as
	\begin{equation}\label{ub}
		E{E_\mathrm{ub}} = \frac{{{{\log }_2}\left( {1 + \frac{{{P_\mathrm{t}}\left\| {\mathbf{h}\left( {\bar x} \right)} \right\|_2^2}}{{{\sigma ^2}}}} \right)}}{{{P_\mathrm{t}}}},
	\end{equation}
	if the initial MA position satisfies $x^0=\bar x$, where $\bar x = \mathop {\arg \max }\nolimits_{x \in \mathcal{A}} \left( {\left\| {\mathbf{h}\left( x \right)} \right\|_2^2} \right)$.
\end{theorem}
\begin{IEEEproof}
	Please refer to Appendix \ref{appendixA}.
\end{IEEEproof}

In the following, we focus on the MA position optimization to maximize the user's energy efficiency. For the convenience of subsequent analysis, denoting the entry in the $l$-th row and $n$-th column of $\mathbf{G}$ as ${{g_{l,n}}}$, the closed-form expression of $\left\| \mathbf{h}\left(x \right) \right\|_2^2$ is given by
\begin{align}
	\left\| \mathbf{h}\left(x \right) \right\|_2^2 & = {\mathbf{f}^H}\left( x \right)\mathbf{G}{\mathbf{G}^H}\mathbf{f}\left( x \right) \nonumber\\
	&= X + \sum\limits_{a = 1}^{L - 1} {\sum\limits_{b = a + 1}^L {2{{\left| {{Y_{ab}}} \right|}}\cos \left( {\frac{{2\pi}}{\lambda }x{\vartheta _{ab}} + \angle  {{Y_{ab}}} } \right)} } ,
\end{align}
where $X = \sum\nolimits_{l = 1}^L {\sum\nolimits_{n = 1}^N {{{\left| {{g_{l,n}}} \right|}^2}} }$, ${Y_{ab}} = \sum\nolimits_{n = 1}^N {{g_{a,n}}g_{b,n}^*} $, and ${\vartheta _{ab}} = {\vartheta _b} - {\vartheta _a}$. Although problem \eqref{max_single1} is significantly simplified compared to problem \eqref{max_single}, it remains non-convex due to the objective function and constraint C4 not being concave with respect to $x$. To tackle this issue, we propose an efficient algorithm to optimize the MA position using the Dinkelbach \cite{din} and SCA \cite{sca} methods.

By introducing Dinkelbach variable $\alpha$ and slack variables $\left\{ {\beta ,\gamma ,\delta } \right\}$, problem \eqref{max_single1} can be reformulated as
\begin{align}\label{max_single2}
	& \mathop {\mathrm{max} }\limits_{x,\beta ,\gamma ,\delta} \quad T{\log _2}\left( {1 + \frac{\beta }{{{\sigma ^2}}}} \right) - \frac{{\delta \gamma }}{v} - \frac{\delta }{v}\left( {\alpha P - \alpha {P_\mathrm{t}}} \right) \\
	&\mathrm{s.t.} \quad \mathrm{C3}, \nonumber\\
	&\hspace{2.3em} \mathrm{C5}:T{\log _2}\left( {1 + \frac{\beta }{{{\sigma ^2}}}} \right) - \frac{{\delta \gamma }}{v} \ge R_\mathrm{TH},\nonumber\\
	&\hspace{2.3em} \mathrm{C6}:\beta  \le h\left( x \right),\nonumber\\
	&\hspace{2.3em} \mathrm{C7}:\gamma  \ge {\log _2}\left( {1 + \frac{{h\left( x \right)}}{{{\sigma ^2}}}} \right),\nonumber\\
	&\hspace{2.3em} \mathrm{C8}:\delta  \ge x - {x^0},\nonumber\\
	&\hspace{2.3em} \mathrm{C9}:\delta  \ge {x^0}-x,\nonumber
\end{align}
where $P=E_0v$ is the driver’s power consumption\footnote{Here, we assume that $P\ge P_\mathrm{t}$ to investigate the effect of high movement-related power on the energy efficiency of the MA system.}. $\alpha$ is given and can be iteratively updated as \cite{din}
\begin{equation}\label{alpha}
	\alpha^i=\frac{{\left( {T - \frac{{\left| {x^i - {x^0}} \right|}}{v}} \right){{\log }_2}\left( {1 + \frac{{{P_\mathrm{t}}\left\| \mathbf{h}\left(x^i \right) \right\|_2^2}}{{{\sigma ^2}}}} \right)}}{{E_0\left| {x^i - {x^0}} \right| + {P_\mathrm{t}}\left( {T - \frac{{\left| {x^i - {x^0}} \right|}}{v}} \right)}},
\end{equation}
where $i$ is the iteration index and $x^i$ is the MA position in the $i$-th iteration. $h\left(x \right) $ is defined as
\begin{equation}\label{hx}
	h\left(x \right)=P_\mathrm{t}X + \sum\limits_{a = 1}^{L - 1} {\sum\limits_{b = a + 1}^L {2P_\mathrm{t}{{\left| {{Y_{ab}}} \right|}}\cos \left( {\frac{{2\pi}}{\lambda }x{\vartheta _{ab}} + \angle  {{Y_{ab}}} } \right)} }.
\end{equation}
However, the resulting term $\delta \gamma$ in the objective function and constraint C5, and $h\left(x \right) $ in the constraints C6 and C7 are neither convex nor concave. Thus, with given local points $\left\{ {{\delta ^i},{\gamma ^i}} \right\}$, a convex surrogate function, which serves as an upper bound for the term $\delta \gamma$, can be constructed as \cite{sca}
\begin{equation}
	\delta \gamma  \le \frac{1}{2}\left( {\frac{{{\gamma ^i}}}{{{\delta ^i}}}{\delta ^2} + \frac{{{\delta ^i}}}{{{\gamma ^i}}}{\gamma ^2}} \right). 
\end{equation}
Then, the lower bound of the objective function in problem \eqref{max_single2} is given by
\begin{equation}\label{obj}
	T{\log _2}\left( {1 + \frac{\beta }{{{\sigma ^2}}}} \right) -\frac{1}{2}\left( {\frac{{{\gamma ^i}}}{{{\delta ^i}}}{\delta ^2} + \frac{{{\delta ^i}}}{{{\gamma ^i}}}{\gamma ^2}} \right) - \frac{\delta }{v}\left( {\alpha P - \alpha {P_\mathrm{t}}} \right),
\end{equation}
which is concave with respect to $\left\{\beta ,\gamma ,\delta\right\}$. Besides, constraint C5 also becomes concave with respect to $\left\{\beta ,\gamma ,\delta\right\}$, i.e.,
\begin{equation}
	\mathrm{C10}: T{\log _2}\left( {1 + \frac{\beta }{{{\sigma ^2}}}} \right) - \frac{1}{2}\left( {\frac{{{\gamma ^i}}}{{{\delta ^i}}}{\delta ^2} + \frac{{{\delta ^i}}}{{{\gamma ^i}}}{\gamma ^2}} \right) \ge R_\mathrm{TH}.
\end{equation}
In addition, for constraints C6, a concave function serving as a lower bound for $h\left( x\right) $ can be obtained by applying the second-order Taylor expansion \cite{sca}, i.e, 
\begin{align}
	h\left( x \right) & \ge h\left( {{x^i}} \right) \nonumber\\
	&+ \frac{{\mathrm{d}h\left( {{x^i}} \right)}}{{\mathrm{d}x}}\left( {x - {x^i}} \right) - \frac{{{\varepsilon}}}{2}{\left( {x - {x^i}} \right)^2} = h_\mathrm{lb}^i\left( x \right),
\end{align}
where
\begin{align}
	&\frac{{\mathrm{d}h\left( {{x^i}} \right)}}{{\mathrm{d}x}} \nonumber\\
	&= \sum\limits_{a = 1}^{L - 1} {\sum\limits_{b = a + 1}^L { - \frac{{4\pi{P_\mathrm{t}}\left| {{Y_{ab}}} \right|{\vartheta _{ab}}}}{\lambda }\sin \left( {\frac{{2\pi}}{\lambda }{x^i}{\vartheta _{ab}} + \angle  {{Y_{ab}}} } \right)} } ,
\end{align}
and $\varepsilon$ is a positive real number satisfying ${\varepsilon} \ge \frac{{{\mathrm{d}^2}h\left( x \right)}}{{\mathrm{d}{x^2}}}$, with the closed-form expression given in Appendix \ref{appendixB}. As a result, constraint C6 can be restated as
\begin{equation}
	\mathrm{C11}: \beta \le h_\mathrm{lb}^i\left( x \right).
\end{equation}
Furthermore, constraint C7 can be rewritten as
\begin{equation}
	{\sigma ^2}{2^\gamma } - {\sigma ^2} \ge h\left( x \right),
\end{equation}
where ${\sigma ^2}{2^\gamma } - {\sigma ^2}$ is lower-bounded by its first-order Taylor expansion, i.e., 
\begin{equation}
	{\sigma ^2}{2^\gamma } - {\sigma ^2} \ge {\sigma ^2}{2^{{\gamma ^i}}} - {\sigma ^2} + {\sigma ^2}{2^{{\gamma ^i}}}\left( {\gamma  - {\gamma ^i}} \right)\ln 2.
\end{equation}
Moreover, a convex upper bound for $h\left( x\right) $ can be constructed as \cite{sca}
\begin{equation}
	h_\mathrm{ub}^i\left( x \right)=h\left( {{x^i}} \right) + \frac{{\mathrm{d}h\left( {{x^i}} \right)}}{{\mathrm{d}x}}\left( {x - {x^i}} \right) + \frac{\varepsilon }{2}{\left( {x - {x^i}} \right)^2},
\end{equation}
Hence, constraint C7 can be restated as
\begin{equation}
	\mathrm{C12}: {\sigma ^2}{2^{{\gamma ^i}}} - {\sigma ^2} + {\sigma ^2}{2^{{\gamma ^i}}}\left( {\gamma  - {\gamma ^i}} \right)\ln 2 \ge h_\mathrm{ub}^i\left( x \right).
\end{equation}
As such, problem \eqref{max_single2} can be transformed into the following convex optimization problem:
\begin{align}\label{max_single3}
	& \mathop {\mathrm{max} }\limits_{x,\beta ,\gamma ,\delta} \quad \eqref{obj} \\
	&\mathrm{s.t.} \quad \mathrm{C3}, \mathrm{C8},\mathrm{C9},\mathrm{C10},\mathrm{C11},\mathrm{C12}, \nonumber
\end{align}
which can be solved optimally by CVX. The proposed algorithm for solving problem \eqref{max_single3} is summarized in Algorithm \ref{alg}.
\begin{algorithm}[!t]
	\caption{Proposed algorithm for solving problem \eqref{max_single3}}
	\label{alg}
	%	\small
	\renewcommand{\algorithmicrequire}{\textbf{Initialization:}}
	\renewcommand{\algorithmicensure}{\textbf{Output:}}
	\begin{algorithmic}[1]
		\REQUIRE Set initial values $\left\{{x^0,\beta^0 ,\gamma^0 ,\delta^0}\right\}$, iteration index $i = 0$, and error tolerance $0 \le \epsilon \ll 1$.
		\ENSURE The optimized MA position $x$.
		\STATE Calculate initial objective value \eqref{obj} and Dinkelbach variable $\alpha^0$ by \eqref{alpha};
		\REPEAT
		\STATE Set $i=i+1$;
		\STATE Solve problem \eqref{max_single3} for the given $\alpha^{i-1}$ and store the intermediate solution $\left\{{x^i,\beta^i ,\gamma^i ,\delta^i}\right\}$;	
		\STATE Update $\alpha^i$ by \eqref{alpha};	
		\UNTIL {Increase of objective value \eqref{obj} is less than $\epsilon$}
		\RETURN $x=x^i$.
	\end{algorithmic}
\end{algorithm}
\section{Simulation Result}
\begin{table}[!t]
	\renewcommand{\arraystretch}{1.5}
	\caption{Simulation Parameters}
	\label{tab1}
	\centering
	\begin{tabular}{lll}
		\toprule
		\textbf{Parameter} & \textbf{Description} & \textbf{Value} \\
		\midrule
		$\lambda$ & Carrier wavelength & 0.01m \\
		$A$ & Length of moving region & 0.02m \\
		$N$ & Number of antennas at the BS & 16 \\
		$L$ & Number of channel paths & 10 \\
		$\rho_0$ & Path loss at the reference distance & $-40$dB \\
		$d$ & Distance between the BS and the user & 50m \\
		$\tilde \alpha$ & Path loss exponent & 2.8 \\
		$\epsilon$ & Error tolerance & $10^{-4}$ \\
		$P_{\mathrm{t}}$ & Maximum transmit power & 10dBm \\
		$P$ & Power consumption of MA movement & 0.5W \\
		$v$ & MA moving speed & 0.2m/s \\
		$T$ & Transmission block duration & 5s \\
		$R_\mathrm{TH}$ & Minimum throughput requirement & 5bits/Hz \\
		$\sigma^2$ & Average noise power & $-70$dBm \\
		\bottomrule
	\end{tabular}
\end{table}
%\begin{table}[!t]
%	\caption{Simulation Parameters}
%	\label{tab1}
%	\centering
%	\begin{tabular}{|l|l|l|}
%		\hline
%		\multicolumn{1}{|c|}{\textbf{Parameter}} & \multicolumn{1}{c|}{\textbf{Description}} & \multicolumn{1}{c|}{\textbf{Value}} \\ \hline
%		$\lambda$ & Carrier wavelength & 0.01m \\ \hline
%		$A$ & Length of moving region & 0.02m \\ \hline
%		$N$ & Number of antennas at the BS & 16 \\ \hline
%		$L$ & Number of channel paths & 10 \\ \hline
%		$\rho_0$ & Path loss at the reference distance & -40dB \\ \hline
%		$d$ & Distance between the BS and the user & 50m \\ \hline
%		$\tilde \alpha$ & Path loss exponent & 2.8 \\ \hline
%		$\epsilon$ & Error tolerance  & $10^{-4}$   \\ \hline
%		$P_{\mathrm{t}}$ & Maximum transmit power & 10dBm   \\ \hline
%		$P$ & Power consumption of MA movement & 0.5W   \\ \hline
%		$v$ & MA moving speed & 0.2m/s   \\ \hline
%		$T$ & Transmission block duration & 5s   \\ \hline
%		$R_\mathrm{TH}$ & Minimum throughput requirement & 5bits/Hz   \\ \hline
%		$\sigma^2$ & Average noise power & -70dBm   \\ \hline
%	\end{tabular}
%\end{table}
This section provides simulation results to evaluate the performance of the proposed scheme. In the simulation, we adopt the channel model in \eqref{channel}, where each element in $\mathbf{G}$ follows circularly symmetric complex Gaussian distribution $\mathcal{CN}\left(0,\frac{\rho_0 d^{-\tilde \alpha }}{L} \right)$. $\rho_0$ denotes the path loss at the reference distance of 1m. $d$ is the distance between the BS and the user. $\tilde \alpha$ represents the path loss exponent. The initial MA position is defined as the center of the moving region, i.e., $x^0=\frac{A}{2}$. The elevation and azimuth AoAs of the channel paths, $\theta_l$ and $\phi _{l}$, are assumed to be independent and identically distributed, with the uniform distribution over the range $\left[0,\pi \right] $. Unless otherwise specified, the default simulation parameters are listed in Table\;\ref{tab1} based on the typical parameters of commercial stepper motors \cite{driver}. 

The proposed scheme is labeled by ``Proposed''. Besides, four benchmark schemes are defined for performance comparison. 1) \textbf{Upper bound}: The upper bound is determined using \eqref{ub}, where $\bar x = \mathop {\arg \max }\nolimits_{x \in \mathcal{A}} \left( {\left\| {\mathbf{h}\left( x \right)} \right\|_2^2} \right)$. 2) \textbf{Max throughput}: The MA position is optimized with the objective of maximizing the achievable throughput, as given in \eqref{R_k}. 3) \textbf{Max SNR}: The MA position is optimized with the objective of maximizing the user's SNR, as given in \eqref{SNR}. 4) \textbf{FPA}: The user is equipped with a single FPA.

\begin{figure}[!t]
	\centering
	\includegraphics[width=1\linewidth]{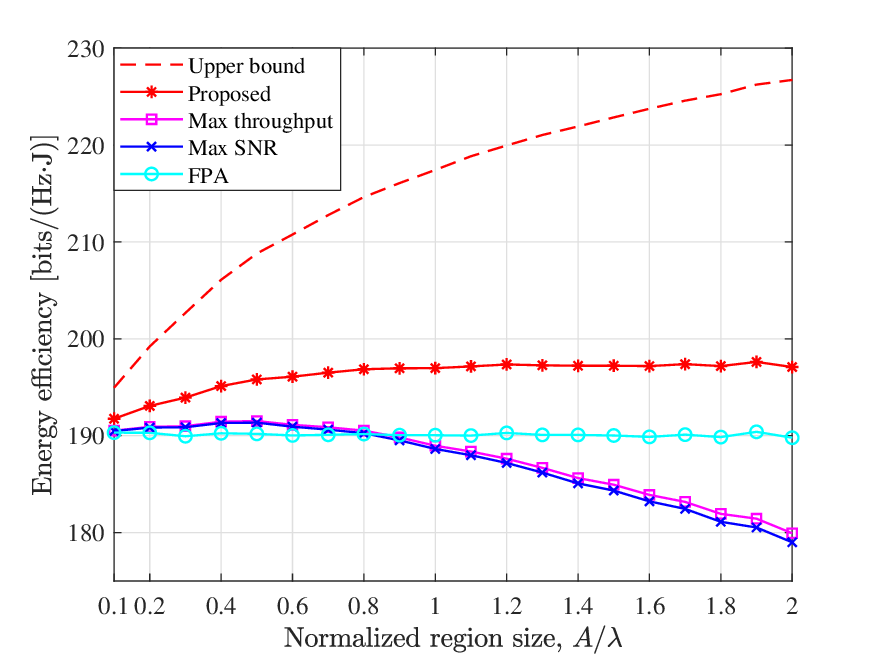}
	\caption{Energy efficiency versus normalized region size.}
	\label{A}
\end{figure}
Fig. \ref{A} shows the energy efficiency of different schemes versus the normalized region size. We can see that the upper bound increases with the enlargement of the moving region. This is because 1) a larger moving region allows the MA to better exploit the spatial DoFs, thereby identifying positions with higher channel gain; and 2) the upper bound in \eqref{ub} does not include any movement-related energy and time consumption. Similarly, the energy efficiency of the proposed scheme also increases with the expansion of the moving region, and it outperforms the conventional FPA scheme. This indicates that, despite the additional energy and time overheads caused by MA movement, the MA system can still demonstrate an energy efficiency gain over the FPA system via MA position optimization. Moreover, as shown in Max throughput and Max SNR schemes, focusing solely on maximizing achievable throughput or SNR results in a significant decrease in the system's energy efficiency as the moving region increases. This is because a large moving region may cause long-distance MA movement.

\begin{figure}[!t]
	\centering
	\includegraphics[width=1\linewidth]{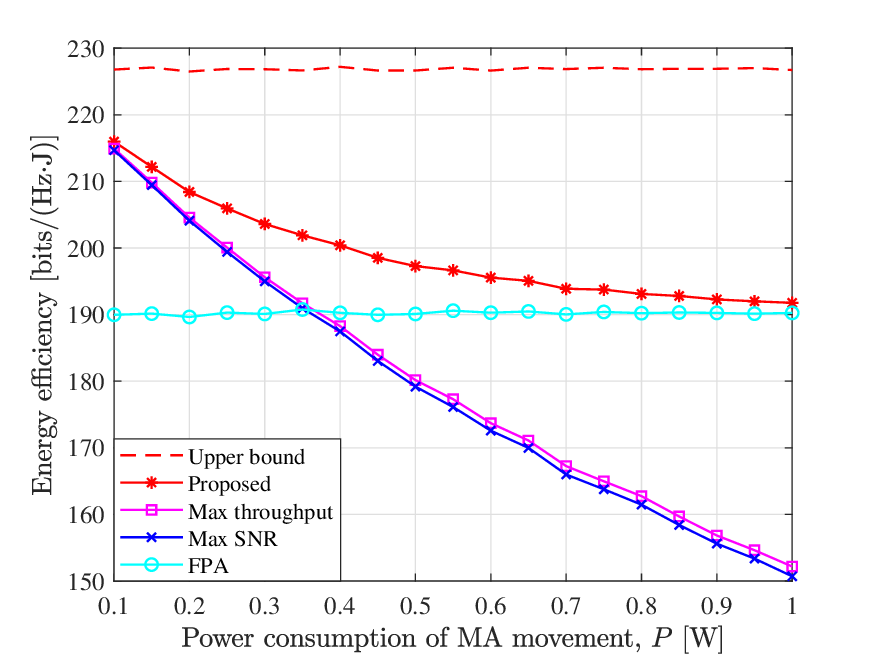}
	\caption{Energy efficiency versus power consumption of MA movement.}
	\label{P}
\end{figure}

Fig. \ref{P} illustrates the energy efficiency versus the power consumption of MA movement for different schemes. Generally, with higher power consumption due to MA movement, the system tends to minimize the MA movement distance to maximize energy efficiency. Therefore, the energy efficiency of the proposed scheme decreases as $P$ increases, eventually converging to the same level as that of the FPA. Furthermore, the energy efficiency of both Max throughput and Max SNR schemes sharply decreases as $P$ increases, because these schemes neglect the energy consumption associated with MA movement and focus solely on improving communication quality. As a result, it is crucial to optimize the MA system's performance by considering both energy consumption and communication quality. In practical applications, drivers with lower power consumption can be employed to control MA movement, thereby yielding higher energy efficiency improvements.
\section{Conclusion}
This paper investigated the energy efficiency optimization for MA systems by modeling the energy consumption caused by MA movement. We derived the upper bound on energy efficiency for the single-user downlink communication system with a single MA based on the MRC beamformer. Moreover, we proposed an effective SCA-based algorithm for optimizing the MA position. The simulation results demonstrated that, although MA movement introduces extra time and energy costs, the MA system achieves an energy efficiency gain compared to the conventional FPA system. The results offer engineering insights for practical applications.
\section*{Acknowledgments}
The calculations were supported by the High-Performance Computing Platform of Peking University.
{\appendices
	\section{Proof of Theorem \ref{theorem1}}\label{appendixA}
	Define $\bar x$ as the position with the maximum channel gain, i.e., $\bar x = \mathop {\arg \max }\nolimits_{x\in \mathcal{A} } \left( {\left\| {\mathbf{h}\left( x \right)} \right\|_2^2} \right)$. Then, we have 
	\begin{align}
		EE\left( x \right) & = \frac{{\left( {T - \frac{{\left| {x - {x^0}} \right|}}{v}} \right){{\log }_2}\left( {1 + \frac{{{P_\mathrm{t}}\left\| {\mathbf{h}\left( x \right)} \right\|_2^2}}{{{\sigma ^2}}}} \right)}}{{E_0\left| {x - {x^0}} \right| + {P_\mathrm{t}}\left( {T - \frac{{\left| {x - {x^0}} \right|}}{v}} \right)}}  \nonumber\\
		&=\frac{{{{\log }_2}\left( {1 + \frac{{{P_\mathrm{t}}\left\| {\mathbf{h}\left( x \right)} \right\|_2^2}}{{{\sigma ^2}}}} \right)}}{{\frac{{E_0\left| {x - {x^0}} \right|}}{{T - \frac{{\left| {x - {x^0}} \right|}}{v}}} + {P_\mathrm{t}}}} \mathop \le \limits^{\left( {{a_1}} \right)} \frac{{{{\log }_2}\left( {1 + \frac{{{P_\mathrm{t}}\left\| {\mathbf{h}\left( x \right)} \right\|_2^2}}{{{\sigma ^2}}}} \right)}}{{{P_\mathrm{t}}}}  \nonumber\\
		&\mathop \le \limits^{\left( {{a_2}} \right)} \frac{{{{\log }_2}\left( {1 + \frac{{{P_\mathrm{t}}\left\| {\mathbf{h}\left( {\bar x} \right)} \right\|_2^2}}{{{\sigma ^2}}}} \right)}}{{{P_\mathrm{t}}}} = E{E_\mathrm{ub}},
	\end{align}
	where the inequality marked by $\left( {{a_1}} \right)$ holds because ${\frac{{E_0\left| {x - {x^0}} \right|}}{{T - \frac{{\left| {x - {x^0}} \right|}}{v}}}} \ge 0$, and the equality can be achieved when $x=x^0$. The inequality marked by $\left( {{a_2}} \right)$ holds because ${\left\| {\mathbf{h}\left( \bar x \right)} \right\|_2^2}$ is the maximum channel gain, and the equality can be achieved when $x=\bar x$. Hence, when $x^0=\bar x$, the upper bound of energy efficiency, $EE_\mathrm{ub}$, can be achieved. This thus completes the proof.
	\section{Construction of $\varepsilon$}\label{appendixB}
	Based on the expression of $h\left(x \right) $ in \eqref{hx}, we have
	\begin{align}
		&\frac{{{\mathrm{d}^2}h\left( x \right)}}{{\mathrm{d}{x^2}}} \nonumber\\
		&= \sum\limits_{a = 1}^{L - 1} {\sum\limits_{b = a + 1}^L { - \frac{{8\pi^2{P_\mathrm{t}}\left| {{Y_{ab}}} \right|\vartheta _{ab}^2}}{{{\lambda ^2}}}\cos \left( {\frac{{2\pi}}{\lambda }x{\vartheta _{ab}} + \angle {Y_{ab}}} \right)} }.
	\end{align}
	Since ${\varepsilon} \ge \frac{{{\mathrm{d}^2}h\left( x \right)}}{{\mathrm{d}{x^2}}}$, we can select $\varepsilon$ as
	\begin{equation}
		\varepsilon=\sum\limits_{a = 1}^{L - 1} {\sum\limits_{b = a + 1}^L {\frac{{8\pi^2{P_\mathrm{t}}\left| {{Y_{ab}}} \right|\vartheta _{ab}^2}}{{{\lambda ^2}}}} }.
	\end{equation}
}


\begin{thebibliography}{1}
	\bibliographystyle{IEEEtran}
	\bibitem{MA1}
	L. Zhu, W. Ma, and R. Zhang, ``Movable antennas for wireless communication: Opportunities and challenges,'' \textit{IEEE Commun. Mag.}, vol. 62, no. 6, pp. 114-120, Jun. 2024.
	
	\bibitem{near1}
	J. Ding, L. Zhu, Z. Zhou, B. Jiao, and R. Zhang, ``Near-field multiuser communications aided by movable antennas,'' \textit{IEEE Wireless Commun. Lett.}, vol. 14, no. 1, pp. 138-142, Jan. 2025.
		
	\bibitem{MA2}
	L. Zhu, W. Ma, and R. Zhang, ``Modeling and performance analysis for movable antenna enabled wireless communications,'' \textit{IEEE Trans. Wireless Commun.}, vol. 23, no. 6, pp. 6234-6250, Jun. 2024.
\newpage
	\bibitem{MAFD1}
	J. Ding, Z. Zhou, W. Li, C. Wang, L. Lin, and B. Jiao, ``Movable antenna-enabled co-frequency co-time full-duplex wireless communication,'' \textit{IEEE Commun. Lett.}, vol. 28, no. 10, pp. 2412-2416, Oct. 2024.
	
	\bibitem{MAFD2}
	J. Ding, Z. Zhou, C. Wang, W. Li, L. Lin, and B. Jiao, ``Secure full-duplex communication via movable antennas,'' in \textit{Proc. IEEE Global Commun. Conf. (GLOBECOM)}, Cape Town, South Africa, Dec. 2024, pp. 885-890.
	
	\bibitem{MAFD3}
	J. Ding, Z. Zhou, and B. Jiao, ``Movable antenna-aided secure full-duplex multi-user communications,'' \textit{IEEE Trans. Wireless Commun.}, vol. 24, no. 3, pp. 2389-2403, Mar. 2025.
	
	\bibitem{MAsate1}
	L. Zhu, X. Pi, W. Ma, Z. Xiao, and R. Zhang, ``Dynamic beam coverage for satellite communications aided by movable-antenna array,'' \textit{IEEE Trans. Wireless Commun.}, vol. 24, no. 3, pp. 1916-1933, Mar. 2025.
	
	\bibitem{MAsate2}
	L. Lin, J. Ding, Z. Zhou, and B. Jiao, ``Power-efficient full-duplex satellite communications aided by movable antennas,'' \textit{IEEE Wireless Commun. Lett.}, vol. 14, no. 3, pp. 656-660, Mar. 2025.
	
	\bibitem{MAISAC1}
	J. Ding, Z. Zhou, X. Shao, B. Jiao, and R. Zhang, ``Movable antenna-aided near-field integrated sensing and communication,'' 2024, \textit{arXiv:2412.19470}.
	
	\bibitem{MAISAC2}
	W. Ma, L. Zhu, and R. Zhang, ``Movable antenna enhanced wireless sensing via antenna position optimization,'' \textit{IEEE Trans. Wireless Commun.}, vol. 23, no. 11, pp. 16575-16589, Nov. 2024.
	
	\bibitem{MAtime}
	H. Wang, Q. Wu, Y. Gao, W. Chen, W. Mei, G. Hu, and L. Xu, ``Throughput maximization for movable antenna systems with movement delay consideration'', 2024, \textit{arXiv:2411.13785}.
	
	\bibitem{MAenergy}
	Y. Wu, D. Xu, D. W. K. Ng, W. Gerstacker, and R. Schober, ``Globally optimal movable antenna-enhanced multi-user communication: Discrete antenna positioning, motion power consumption, and imperfect CSI,'' 2024, \textit{arXiv:2408.15435}.
	
	\bibitem{din}
	W. Dinkelbach, ``On nonlinear fractional programming,'' \textit{Manag. Sci.}, vol. 13, no. 7, pp. 492-498, Mar. 1967. [Online]. Available: https://doi.org/10.1287/mnsc.13.7.492
	
	\bibitem{sca}
	Y. Sun, P. Babu, and D. P. Palomar, ``Majorization-minimization algorithms in signal processing, communications, and machine learning,'' \textit{IEEE Trans. Signal Process.}, vol. 65, no. 3, pp. 794-816, Feb. 2017.
	
	\bibitem{driver}
	Faulhaber, \textit{Stepper Motors, Series AM2224}, 2023. [Online]. Available: https://www.faulhaber.com/fileadmin/Import/Media/EN\_AM2224\_FPS\\.pdf
\end{thebibliography}
\end{document}